\documentstyle[prd,aps,psfig]{revtex}
\begin{document}

\title{Extension of the Geometrical Two-Chain Model to high energies}
\draft
\author{T. Wibig}

\address{Experimental Physics Dept., University of \L odz,
Pomorska 149/153, 90-236 \L odz, Poland}

\date{\today}

\maketitle

\begin{abstract}
In the paper the extension of the Geometrical Two-Chain model to higher
energies is presented. The agreement with experimental data is achieved using
pre-hadronization chain breakups due to coloured dipole radiation
mechanism and the discretization of the soft gluon emission process.
\end{abstract}

\pacs{13.85.-t, 12.39.-x, 13.87.Fh, 12.40.-y}

\section {Introduction}

The Geometrical Two--Chain (G2C) model was established to describe
relatively
low (up to $\sqrt{s} \sim 100$ GeV) multiparticle production processes.
The model gives an unified picture
of wide range of processes: from elastic hadron scattering and
${\rm e^+e^-} \rightarrow {\rm hadrons}$ reactions (Ref.\cite{ws1}),
through diffractive dissociation in hadron-proton interactions
(Ref.\cite{sw}), to non-single-diffraction hadron-hadron
(Ref.\cite{ws2}) and hadron-nucleus interactions (Ref.\cite{ws3}).
The geometrization of the interaction picture was obtained using
the eikonal function formalism in terms of hadronic matter distribution
of interacting hadron (assuming point--like leptons).

The Geometrical Two--Chain model acts in two steps. The first is the soft
scattering which
turns incoming particles into two intermediate objects. They can be,
to some extend,
treated as chains concerning quarks (di--quarks) on their ends similar to
the well known string structures of the LUND (Ref.\ \cite{lund})
or Dual Parton (Ref.\ \cite{dpm}) models.
The second stage of the interaction is the hadronization of outgoing chains.

The creation of chains (strings) and their consecutive fragmentation, in the
very general sense, is a common feature of all contemporary multiparticle
production process descriptions. However, the particular realization
differs from model to model. In the G2C model chain creation is not
controlled by respective structure functions. In that sense it is rather
conservative and old--fashion model but, on the other hand, such treatment has
some advantages. Our knowledge of the structure function behaviour,
specially at low $x$ values, is still not perfect. In our model the
parametrization used does not pretend to be a fundamental law of 
the nature but
it is just {\em parametrization} which can be changed and tuned to new data
as they will appear.
Generally speaking our geometrization strategy is not in contradiction to the
standard structure function approach.

\section {Geometrical Two-Chain Model}

In the G2C model two chains are created on the first stage of the interaction
process. The invariant mass of each of them ($M_{1,2}$) is determined by
the impact parameter $b$ of the particular collision and the CMS available
energy ($\sqrt{s}$):

\begin{equation}
M_{1,2} \ \sim  \ M_0 \: + \:
\left( {{\Omega (b)} \over {\Omega (0)}} \right)
^\alpha
\: \left({ {\sqrt{s} \over 2 } \: -  \: M_0 } \right)  ,
\label{mchain}
\end{equation}

\noindent
where: $M_0$ is a mass of the lightest hadron which can be formed from the
particular quark contents of the chain and $\Omega$ is related to the mass
distributions in the colliding hadrons ($\rho _{1,2}$) by:

\begin{equation}
\Omega(b) \ = \ \int  {\rm d \vec r} \ \rho _1({\rm \vec b}) \ \rho _2(
{\rm \vec b -  \vec r)}
\label{omega}
\end{equation}

The value of $\alpha$ parameter used in the present work is equal to 0.28.

In the case of ${\rm e^+e^-} \rightarrow {\rm hadrons}$ reaction only one
chain is created. Thus $\sqrt{s}/2$ is replaced by $\sqrt{s}$ and the impact
parameter is set to 0 ($\Omega(b) / \Omega(0) \equiv$ 1).
One chain is created also in single diffractive case and its
mass is chosen randomly from $1/M^2$ distribution. Two such chains appear
in double diffractive events. (In the present paper only the NSD events will
be discussed.)

The hadronization starts later when all chains are formed. This is
specially important for h-{$\cal N$} interactions and for the high energy
interactions in the multi--chain model as it will be discussed below.
Hadrons are created from $q$--$\overline q$ (or diquark--antidiquark) pairs
emerged from the chain CMS energy uniformly in the phase space. This is an
important point of the model: no additional dynamical constrains
are introduced during the hadronization phase but these originated from
kinematical restrictions enforced by momentum and energy conservation laws.
Thus the G2C model can be considered as a minimal
model to study divergences between reality (experiments) and 
non-reducible kinematic (phase space) background.
Such differences can be interpreted as introduced by some dynamics 
of the multiparticle production process.

The transverse momentum of created quarks originate due to tunnelling
mechanism described by the formula:

\begin{equation}
f(p_{\bot})\ \sim \ \exp{\left( - { {m_{\bot}^2 \over \kappa}}\right)}
\label{fpt}
\end{equation}

\noindent
and it is conserved locally. The value of $\kappa$ used in our calculations for
light quark pair creation is equal to 0.28 while for the strange quarks
and diquarks it is 0.38 and 0.6 respectively.
Flavours of crated hadrons are given by the
very few conventional suppression factors like $s/ud$, $qq/q$, $vector/scalar$.
The mass difference of the hadron under consideration and the lightest hadron
which can be build from particular quark configuration is also taken into
account in the way similar to the one given in the Eq.(\ref{fpt}).

The phase space density of the produced hadrons is again parametrized using
the geometrical picture.
The mean number of hadronization breakups of the chain $<n_{q\overline q}>$
($q$--$\overline q$ or $qq$--$\overline {qq}$ pairs) is given by:

\begin {equation}
<n_{q\overline q}>\ = \ n_0 \: \left( \Omega(b) \over \Omega(0) \right) ^\beta
\ ,
\label{nqq}
\end {equation}

\noindent
where:

\begin {equation}
n_0 \: = \ A \: \ln ( M_{\rm chain}) \: + \: B \ .
\label{nqq2}
\end {equation}

The actual number of created pairs $n_{q\overline q}$ is distributed 
due to the poissonian
distribution with the respective mean value. It is
directly related to the number of the first rank hadrons produced in
the event. Values of $\beta$, $A$ and $B$ used are 0.4, 6.7 and -- 6.2
respectively.
The correction for low ($\leq 10$ GeV) chain masses is introduced. The smooth
change of $A$ to the value of 3.0 at $\leq 3$ GeV 
($B$ has to be changed respectively) gives
a reasonable limit for particle multiplicities at extremely low energies 
and can be understand, to some extent, as an effect of the resonant
hadron production.

The created hadron flavours are ordered in rapidity to preserve the
information of the incoming hadron quark contents. For barions
the pop-corn mechanism from the LUND model (Ref.\cite{pop}) was adopted.

\section {Geometrical Multi-Chain Model}

The G2C model as described above reproduces very well a number of
interaction characteristics like e.g. mean multiplicities, multiplicity
distributions, main inclusive distributions of produced particles. It has
been shown in the Refs.\cite{ws1,ws2} up to $\sqrt{s} \sim$ 30 GeV. For
the higher energies, of course, this model has to fail.
Since ISR data has been published the increase of the height of plateau in
inclusive rapidity distributions (what leads to the faster than $\ln (s)$
growth of the mean multiplicity) is a very well experimentally established 
fact. This was one of the most apparent reasons for investigations of a
new physical mechanism responsible for very high energy interaction picture.
In majority of up-to-date models the idea which can be called
multi--string intermediate state is present. One way of introducing it
is developed in DPM--like models (Ref.\ \cite{dpm})
using multi--pomeron structures, another
by using the concept of gluon emission from expanding strings is adapted to
the LUND--class of models (Ref.\ \cite{lund,ari}).

In the framework of geometrical interaction picture 
there is a natural place for
this second approach. Just after formation of two chains in the ''soft
scattering'' phase they can be forced to break up through the ''soft gluon
emission'' before exact hadronization starts. Such a picture is a gist of the
Geometrical Multi--Chain model (GMC) which is the subject of this paper.

\section{Soft Gluon Emission.}

The internal structure of the geometrical chain is in fact quite similar to the
one of the LUND string. There are two coloured objects of relatively
low masses each carrying invariant mass of $M_{\rm chain}$. In the
quark-antiquark (quark-diquark) rest system both coloured ends has to move with
almost the velocity of light so due the existence of the colour charge
the colour dipole radiation has to occur.
Thus, it is expected that the soft gluon brehmsstrahlung
cascade originate in such system. Its mechanism is described
extensively in the ARIADNE code (Ref.\ \cite{ari}) adopted by general
LUND interaction programs. The ARIADNE 4.07 is a particular realization of
this idea and it was used in GMC calculations in the present paper.

The probability of the soft gluon emission is well--known (e.g.
Refs.\cite{glubre,lund}) and can be described with analogy to 
QED using Sudakov form factor in the way:

\begin{equation}
{{d P(p_{\bot}^2,y)} \over {d p_{\bot}^2 d y} } \ = \
{{d \sigma(p_\bot^2,y)} \over {d p_\bot^2 d y }} \
exp{
\left(  - \int _{p_\bot^2}^{{p_\bot^2}_{max}} d^2 k \ {\cal I}
(k^2) \right)}  ,
\label{suda}
\end{equation}

\noindent
where:

\begin{equation}
{\cal I}(k^2) \  = \  \int _{ y_{min} (k^2) }^
{ y_{max} (k^2)} \:
d y' \: {{d \sigma(k^2,y')} \over {d k^2 d y'}} ,
\end{equation}

\begin{equation}
d \sigma \ \sim \ \alpha_s \ {{d p_{\bot}^2} \over {p_{\bot}^2}} \: d y  \ .
\end{equation}

Values of the model parameter used in our GMC realization are as the
default setting in the ARIADNE 4.07 program.
Small difference is connected with the gluon emission suppression related
to a space extension of the emission source.
This works for diquark chain ends only so it is not
important for the ${\rm e^+e^- \rightarrow}$ hadrons processes.
The parameter interpreted as a transverse source size
is set in the present calculations to 1 GeV while in the default ARIADNE
its value is equal to 0.6 GeV. This change is related to the analysis
of Bose--Einstein correlations in hadronic interactions presented in
Ref.\ \cite{be}. However, it is not crucial for the results of
this work. Very similar results could also be obtained with the
value of 0.6 GeV and respective slight change of other model parameters.

The important difference of the GMC model in comparison with standard
ARIADNE soft gluon emission is in the implementation of the discretization
of the process. The idea is described in the Ref.\ \cite{samu}.
In this paper the (logarithmic) phase space (ln(${\rm p}_\bot^2$), y)
is divided into discrete cells and the point is that each cell can be
occupied by only one brehmsstrahlung gluon. The theoretical basis of this 
picture
is widely discussed in the original paper. In the most general way it can be
expressed as following: if there are two gluons occasionally emitted too close
one to the other they combine together conduced finally to one
''effective gluon''. The measure of the closeness can be derived, to
some extend, from the running coupling constant QCD and for two gluons
the minimum distance in rapidity should be of order of 11/6 
(Ref.\ \cite{samu}).

In our model this idea is adopted in the way that gluons are produced by the
standard ARIADNE coloured antenna and then the gluon recombination
initiates. The distance between gluons is defined in three--dimensional
(ln(${\rm p}_{\rm x}^2$), ln(${\rm p}_{\rm y}^2$), y)
space. The recombination procedure starts with the pair of gluons closest
each other and finished when all ''effective gluons'' stay apart by at least
the value of a parameter denoted by $\delta {\rm y_g}$ which is going to be
adjusted to the data.

After soft gluon emissions the main chain is broken into smaller parts. In
the present version of the GMC model each gluon is turned to $q\overline
q$ pair and its momentum is shared between them by halves.

Further, chain pieces (each one containing again $q$--$\overline q$
($qq$--$q$) pair) of smaller masses and with 
some additional transverse momenta
hadronize according to the G2C model described above.

\section{Results and discussion}

If the mean particle multiplicity produced during the hadronization phase is
proportional to the logarithm of the chain mass then the
emission of gluons with non--zero transverse masses and consecutive
breakup of the chain before hadronization leads to the increase of the average
multiplicity. Also the additional transverse momentum originated from the soft
gluon brehmsstrahlung should increase the mean secondary
hadron ${\rm p}_\bot$. The question is: is it possible to reach the
quantitative agreement with measurements preserving the idea described above
with acceptable parameter values?

\begin{figure}
\centerline{\psfig{file=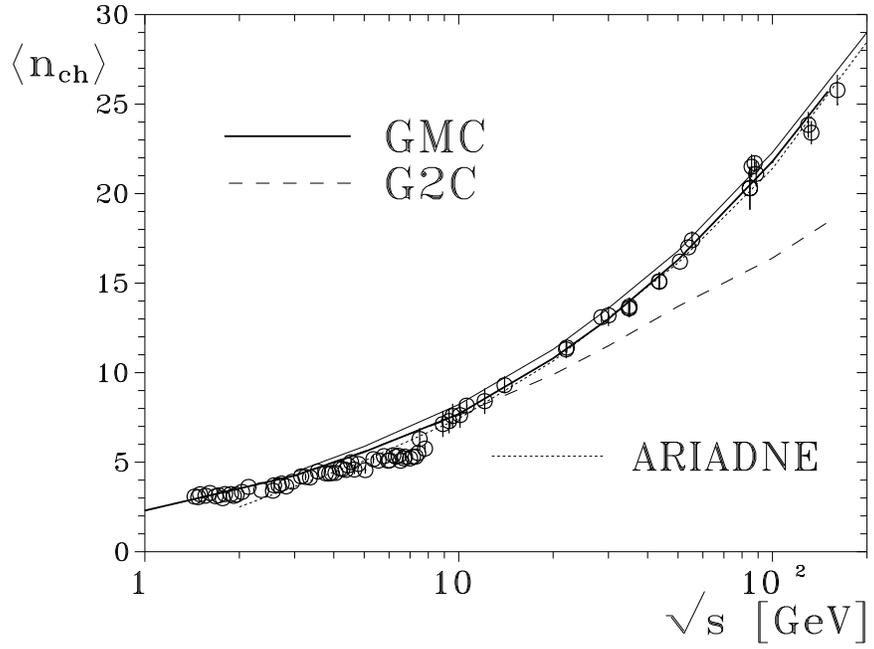,width=10cm}}
\caption{Mean charged multiplicity in ${\rm e^+e^-} \rightarrow {\rm
hadrons}$ reactions. 
The thick solid line represents result of the Geometrical Multi--Chain
model calculations while the long dashed one is for the same parameter value 
set but without soft gluon emission process (pure two--chain picture). 
Thin solid line represents the GMC model results with weaker gluon clustering
($\delta {\rm y_g} = 0$).
The short dashed line is a result of the LUND interaction model calculations.
The data point are from [15].
}
\label{nee}
\end{figure}

As it was done in Ref.\ \cite{ws1} first we proceed with the mean
hadron multiplicities in ${\rm e^+e^-} \rightarrow {\rm hadrons}$ reactions.
The geometrical model parameters $\alpha$ and $\beta$ in Eqs.(\ref{mchain})
and (\ref{nqq}) do not interfere this ($\Omega \equiv
\Omega(0)$). The same can be said about parameters of the suppression of
short wave length gluon emissions by extended (diquark) chain ends in the
ARIADNE model as described in Ref.\ \cite{ari}.
Values of $A$ and $B$ in Eq.(\ref{nqq2}) were first obtained at this point.
The parameter $\delta {\rm y_g}$ of the discrete QCD approximation used has
a small influence on the ${\rm e^+e^-}$ mean multiplicity data due to
relatively small range of chain masses under study (the results for
$\delta {\rm y_g} = 0$ are shown in the Fig.\ref{nee}
by the thin solid line).  
The result of the GMC model calculations 
is presented in the Fig.\ref{nee} by the thick solid line. It was 
obtained with the $\delta {\rm y_g} = 1.45$ adjusted using mainly other 
data what will be shown below. Results of
calculations without the soft gluon emission (pure G2C model with the same
$A$ and $B$ parameter values) are given by the dashed line. Expected
enhancement is clearly seen and the accuracy achieved is very satisfactory.

Predictions of the LUND interaction model (ARIADNE + JETSET with the
set of parameters adjusted to the DELPHI data) are given also
by the short dashed line. At this point it should be clarified why the LUND
model leads to the very similar results as ours which uses also ARIADNE
procedures to describe soft gluon emissions but with the discretization
procedure hereafter. There are of course differences in the hadronization
description but they are not essential here. The important difference is
in the treatment of coloured antenna radiated gluons.
The discretization process used in our GMC model works on a {\em gluon}
level. Finally each gluon (''effective gluon'') has to form a
common end of two chains which then fragment independently. Thus each
soft emission creates one additional chain which is forced to
hadronize to at least one final hadron.
In the LUND picture the low mass (below some critical value) {\em jets} 
are combined with the nearby ones before they hadronize.
Final results, as it can be seen, are similar but the underlying 
physics is quite different.

In the next step parameters $\alpha$ and $\beta$ in Eqs. (\ref{mchain}) and
(\ref{nqq}) has to be adjusted using the proton--proton multiplicity data.

\begin{figure}
\centerline{\psfig{file=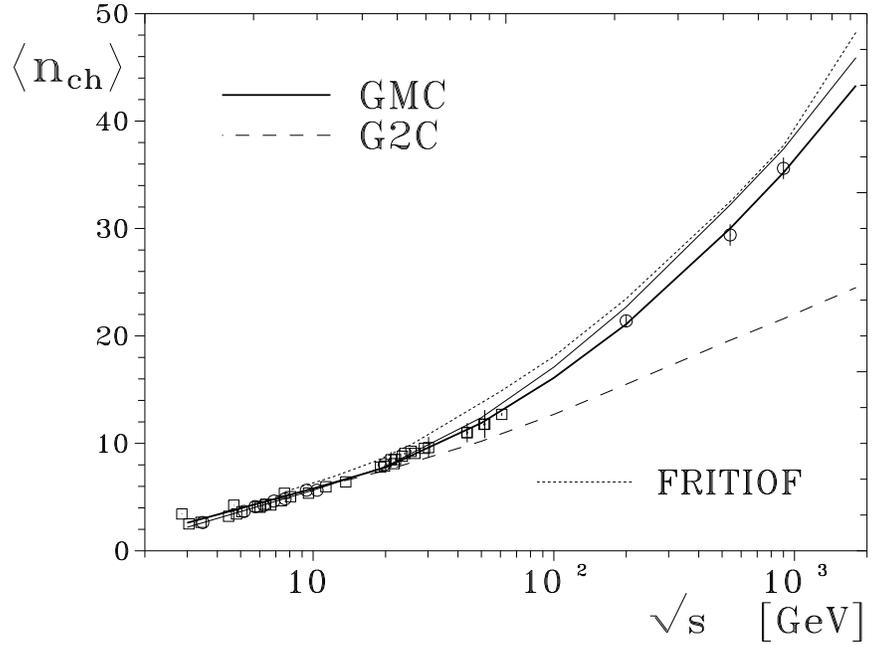,width=10cm}}
\caption{Mean charged multiplicity in hadron--proton multiparticle production
reactions. 
The thick 
solid line represents result of the Geometrical Multi--Chain
(GMC) model calculations while the long dashed one is for the pure two--chain
(G2C) interaction picture. Thin solid line represents the GMC model results 
with weaker gluon clustering ($\delta {\rm y_g} = 0$).
The short dashed line is a result of the LUND 
model calculations.
The data point are from [16].
}
\label{npp}
\end{figure}

\begin{figure}
\centerline{\psfig{file=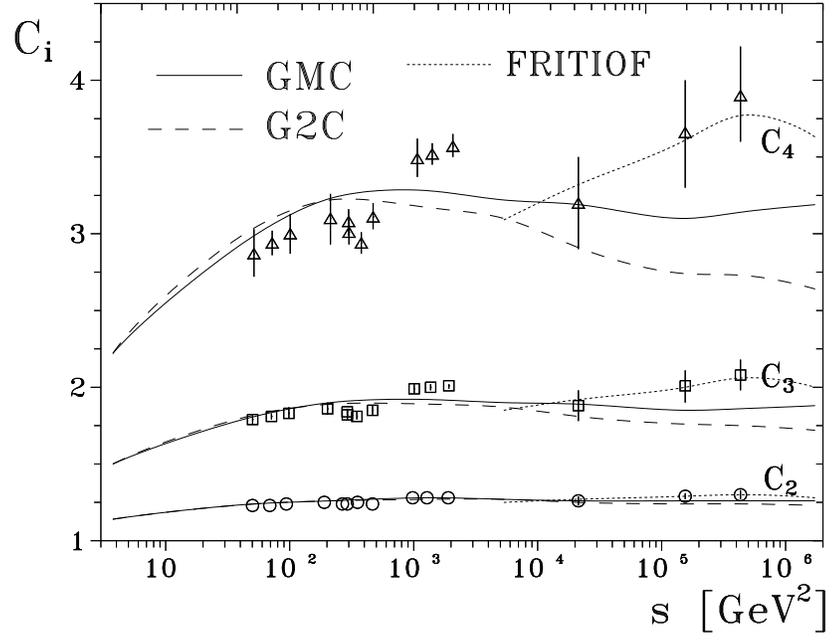,width=10cm}}
\caption{Normalized charged multiplicity distribution moments for hadron--
proton multiparticle production reactions. The  
solid line represents result of the Geometrical Multi--Chain
(GMC) model calculations while the long dashed one is for the pure two--chain
(G2C) interaction picture. 
The short dashed line is a result of the LUND 
model calculations.
The data point are from [16].
}
\label{cpp}
\end{figure}

The mean charged multiplicity, multiplicity moments (defined as
${\rm C_k} = \left \langle {{\rm n_{ch}^{\ k}}} \right \rangle / \left 
\langle {\rm n_{ch}} \right \rangle ^{\rm k}$)
data and the GMC model results are presented in the Figs.\ref{npp}
and \ref{cpp}. The result of G2C calculations is also given.

To see the influence of the discretization of the gluon brehmsstrahlung
process result of the calculation with the $\delta {\rm y_g}$ parameter
equal to 0 is given in the Fig.\ref{nee} and \ref{npp} by the thin solid line.
The difference is not very large but it has to be pointed out here that
the number of gluons emitted primarily is limited by the threshold value
of ${\rm p}_\bot$ of emitted gluons used in coloured dipole radiation
mechanism adopted in ARIADNE code. Its value was set to 0.6 GeV/c.
In principle, such limit is equivalent to the assumption that all softer
gluons are effectively included in harder emissions or hadronization
processes. Thus $\delta {\rm y_g} = 0$ does not mean absence clustering of 
emitted gluons, but only an omission of joining a relatively ''massive'' 
gluons.

The results of the LUND model calculations with the FRITIOF (Ref.\ \cite
{frit}) program are also presented in the Fig.\ref{npp}
by the short dashed line. For the
soft gluon emission there are few different sets of ARIADNE program
parameters. Here the same set called ''DELPHI'' was used as for the 
description of ${\rm e^+e^-} \rightarrow {\rm hadrons}$ mean multiplicities
in the Fig.\ref{nee}.

As it can be seen mean multiplicities are reproduced by GMC model 
very well while
obtained multiplicity distributions are slightly narrower than these observed
in experiments at the highest energies (Fig.\ref{cpp}). 
However, the difference is not very
significant and, what is more important, the tendency of decreasing the higher
moments with the increasing interaction energy seen for G2C converts toward
experimental results.

Obtained values of the model parameters $\alpha$, $\beta$, $A$ and $B$
parameters were then used to determine finally the value of
$\delta {\rm y_g}$ using the data on secondary hadron transverse momenta.

\begin{figure}
\centerline{\psfig{file=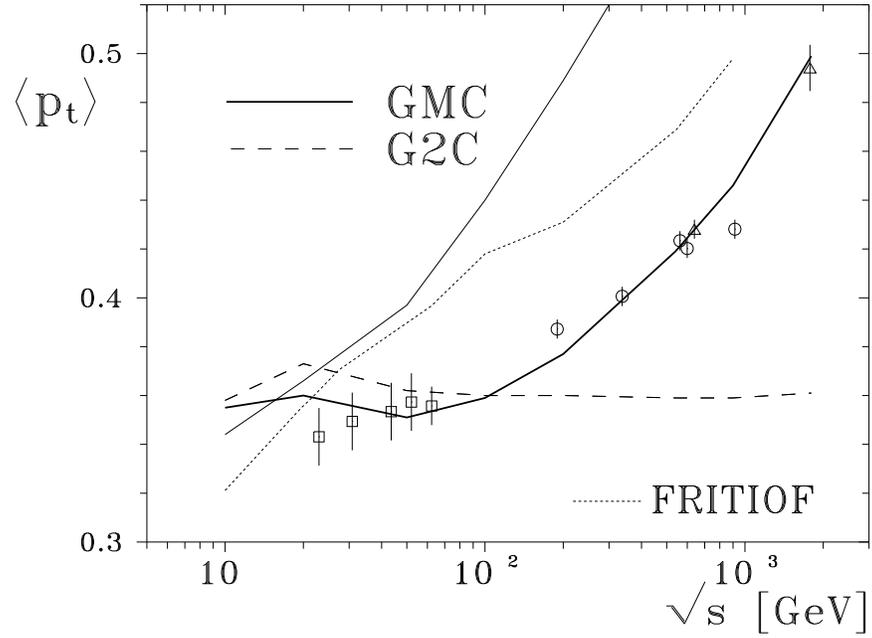,width=10cm}}
\caption{Mean charged particle transverse momentum for hadron--proton
multiparticle production reactions. The line description as in the Fig. 2.
The data point are from [17].
}
\label{ptd}
\end{figure}

The dependence of the averaged transverse momentum on energy is shown
in the Fig.\ \ref{ptd}. The constant value (for high energies) predicted by the
G2C model is settled by the value of the $\kappa$ parameter in the
Eq.(\ref{fpt}). The same value of $\kappa$ with the soft gluon radiation
mechanism gives an increase of average ${\rm p}_\bot$ in perfect agreement
up to the highest available energy accelerator data.
The average transverse momenta of produced particles are quite sensitive 
to the gluon clustering mechanism. 
Thus they were used to determine the value of
$\delta {\rm y_g}$ in the GMC model. For example the very strong 
increase of the average value of ${\rm p}_\bot$ for 
$\delta {\rm y_g} = 0$ is shown in the Fig.\ \ref{ptd} (thin solid line).
Finally $\delta{\rm y_g}$ was found to be equal to 1.45.

The most significant and crucial test of the GMC model is the comparison
with inclusive energy ( longitudinal momentum or (pseudo)rapidity )
data.
The increase of the plateau and the breaking of the Feynman scaling at high
energies are two features of the highest importance in applications of the
model for the hadronic cascade in the thick media.

\begin{figure}
\centerline{\psfig{file=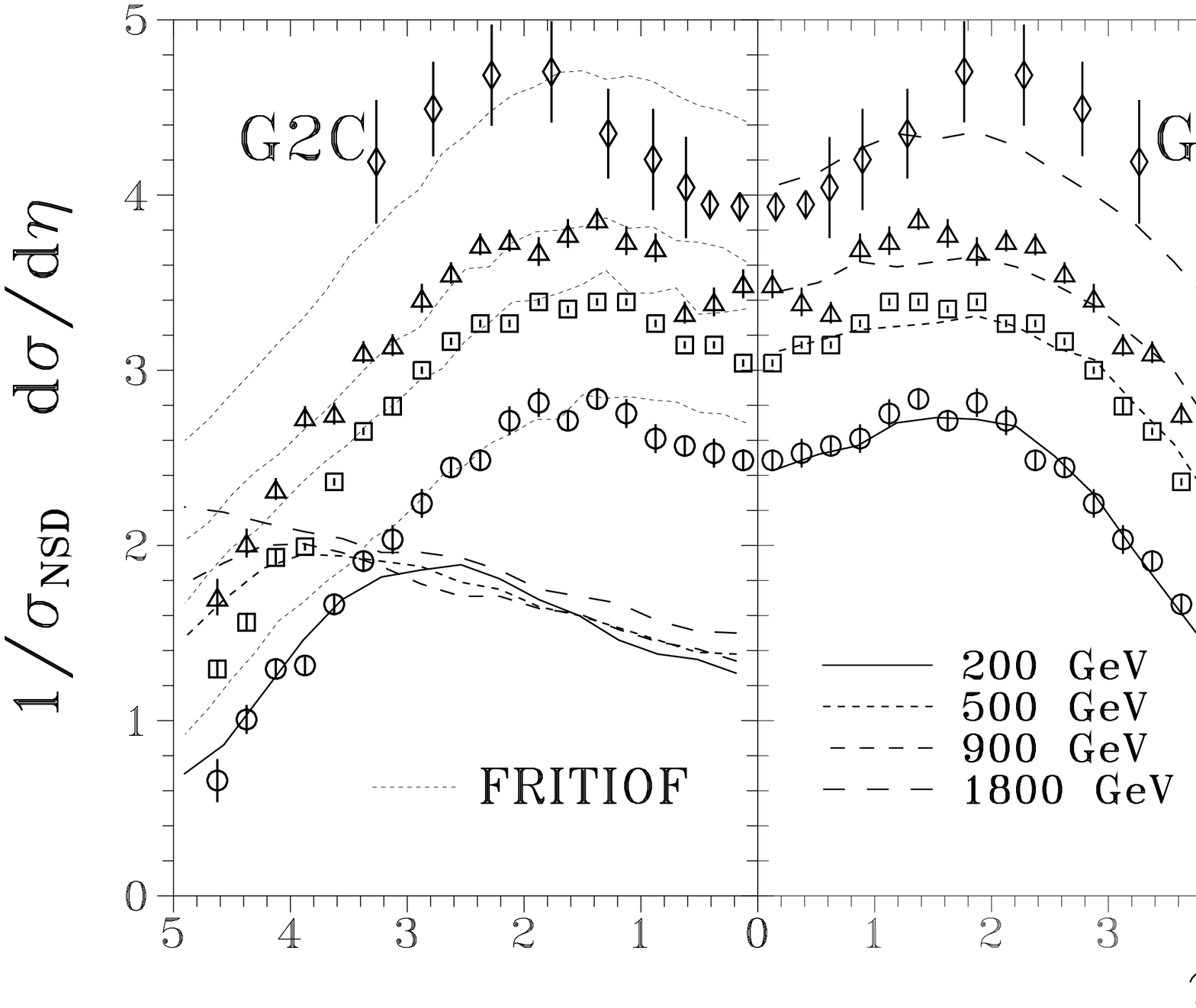,width=10cm}}
\caption{Inclusive rapidity spectra of charged particles in proton--proton
interactions at SPS and Tevatron energies. The results for G2C model are
given in the left part of the Figure while the present GMC model calculations
are in the right part. The short dashed lines in the left part are results of 
the LUND model.
The data point are from [16].
}
\label{y}
\end{figure}

Rapidity distributions for SPS and Tevatron energies are presented in
Fig.\ \ref{y}. The very significant change from G2C to GMC model
predictions is seen and the agreement obtained with the GMC model is
meaningful. The results of FRITIOF with DELPHI set of parameters for 
ARIADNE are also given for comparison.

However good the agreement is, it can not be treated as a proof of the
correctness of the proposed model. There are other very well--known
model available consistent with data as well. Two main classes of such models
are: the one based on DPM picture (There are many particular 
realizations of the DPM idea see e.g. Ref.\ \cite{dpm})
and the second -- relativistic string
model of LUND (Ref.\ \cite{lund}). 
There are many particular realization of the DPM idea.
The most complete comparison between some of them and the data is given in
the Ref.\cite{compa}. The very interesting model calculations are presented
in the Ref.\cite{dpl}. The multi--pomeron exchange concept is combine
there with the impact parameter description and the standard LUND string
fragmentation procedures JETSET are used. The LUND model itself for
proton-proton and ${\rm e^+e^-} \rightarrow {\rm hadrons}$ reactions is
available as a package of FRITIOF, ARIADNE, PYTHIA and JETSET routines.
Results obtained using these codes are presented for all discussed interaction
characteristics in the Figures.
It should be said here that even if the
particular feature is better described by one model it is not
an argument against or for any model.
The LUND interaction picture is of course much more complete what was shown
in a number of papers. The point is that our GMC model with its physical
assumption is {\em also} able to reproduce the data quite well.

\section{Summary}

The extension of the Geometrical Two--Chain multiparticle production
mechanism to the higher energies is obtained by the introduction
of the soft gluon emission process. The framework of the ARIADNE coloured
antenna radiation with the recent idea of discretization of the process
is used to perform high energy chain pre--hadronization breakups. The
default ARIADNE 4.07 parameter values were used. The discrete net step
$\delta{\rm y_g}$ was found to be equal to 1.45 which is close to the
presumed value of 11/6. The difference is not significant according to the
slightly different implementation of the (ln(${\rm p}_\bot^2$), y) cell idea.

The consistency between multiparticle hadronic interactions data and
predictions of the GMC model presented in this paper and the solid 
theoretical basis 
of the model grant a capability of reasonable extrapolation of the
model to high energies.

\begin{references}

\bibitem{ws1}
T. Wibig and D. Sobczy\'{n}ska, Phys. Rev. {\bf D49}, 2268 (1994).

\bibitem{sw}
D. Sobczy\'{n}ska and T. Wibig , Bull. Russ. Acad. Sci. {\bf 58}, 1970
(1994).

\bibitem{ws2}
T. Wibig and D. Sobczy\'{n}ska, Phys. Rev. {\bf D50}, 5657 (1994).

\bibitem{ws3}
T. Wibig and D. Sobczy\'{n}ska, J. Phys.  {\bf G21}, 29 (1995).

\bibitem{lund} B. Anderson, G. Gustafson G. Ingelman and T. Sj\"{o}strand, 
Phys. Rep. {\bf 97}, 31 (1983); H. Bengtsson and T. Sj\"{o}strand, 
Comput. Phys. Commun. {\bf 46}, 43 (1987).

\bibitem{dpm} 
P. Aurenche, F. W. Bopp, A. Capella, J. Kwieci\'{n}ski, M. Maire, J. Ranft 
and J. Tran Thanh Van, Phys. Rev. {\bf D45}, 92 (1992);
K. Werner, Phys. Rep. {\bf 232}, 87 (1993);
A. Capella, U. Sukhatme, C-I Tan and J. Tran Thanh Van,
Phys. Rep. {\bf 236}, 225 (1994);
R. S. Fletcher, T. K. Gaisser, P. Lipari and T. Stanev,
Phys. Rev. {\bf D50}, 5710 (1995);
J. Ranft, Phys. Rev. {\bf D51}, 64 (1995).

\bibitem{pop} B. Anderson, G. Gustafson and T. Sj\"{o}strand, Phys. Scripta
{\bf 32}, 574 (1985).

\bibitem{ari} L. L\"{o}nnblad, Comput. Phus. Commun. {\bf 71}, 15 (1992).

\bibitem{glubre} G. Gustafson, Phys. Lett. B{\bf 175}, 453 (1986); G.
Gustafson and U. Pettersson, Nucl. Phys. B{\bf 306}, 746 (1988); B.
Andersson {\it et al.}, Z. Phys. C{\bf 43}, 625 (1989); B. Andersson, G.
Gustafson and L. L\"{o}nnblad, Nucl. Phys. B{\bf 339}, 393 (1990).

\bibitem{be}
T. Wibig, Phys. Rev. {\bf 54}, 707 (1995).

\bibitem{samu}
B. Andersson, G. Gustafson and J. Samuelsson, Nucl. Phys. B{\bf 463}, 217
(1996).

\bibitem{frit}
B. Andersson, G. Gustafson and Hong Pi, Z. Phys. B{\bf C57}, 485 (1993).

\bibitem{compa}
J. Knapp, D. Heck and G. Schatz, Forschungszentrum Karlsruhe Report
FZKA--5828, Karlsruhe (1996).

\bibitem{dpl}
T. Sj\"{o}strand and M. van Zijl, Phys. Rev. {\bf D36}, 2019 (1987).

\bibitem{nee}
G. Giacomelli, Nucl. Phys. {\bf B25}, 30, (1992);
M. Schmelling, Phys. Scripta {\bf 51}, 683 (1995).

\bibitem{pp}
J. Whitemore, Phys. Rep. {\bf 10}, 273 (1974);
U. Amaldi and K. R. Schubert, Nucl. Phys. {\bf B166}, 301 (1980);
G. J. Alner {\it et al.}, Phys. Lett. {\bf 138B}, 304 (1984);
{\bf 160B}, 199 (1985); 
{\bf 167B}, 476 (1986); 
Phys. Rep., {\bf 154}, 247 (1986); 
Z. Phys. C {\bf 33}, 1 (1986); 
R. E. Ansorge {\it et al.}, Z. Phys. C {\bf 43}, 357 (1989); 
F. Abe {\it et al.}, Phys. Rev. D {\bf 41}, 2330 (1990).

\bibitem{pt}
F. Abe {\it et al.}, Phys. Rev. Lett. {\bf 61}, 1819 (1988).
\end{references}
\end{document}